\documentstyle[aaspp4]{article}

\begin{document}

\title{Temporal and Spectral Correlations of Cyg X-1\footnote{To appear in 
{\sl The Astrophysical Journal}}
}
 
\author{T.P. Li and Y.X. Feng}  
\affil{ High Energy Astrophysics Lab., Institute of High Energy Physics, \\
Chinese Academy of Sciences, Beijing}

\author{L. Chen}  
\affil{Department of Astronomy, Beijing Normal University, Beijing}

\begin{abstract}
Temporal and spectral properties of X-ray rapid variability of Cyg X-1
are studied by an approach of correlation analysis in the time domain 
on different time scales. The correlation coefficients between the total intensity 
in 2--60 keV  and the hardness ratio of 13--60 keV to 2--6 keV band
on the time scale of $\sim 1$ ms are always negative in all states.
For soft states, the correlation coefficients are positive on all the time scales from 
$\sim 0.01$ s to $\sim 100$ s, which is significantly different with
that for transition and low states. Temporal structures in high energy band
are  narrower than that in low energy band in quite a few cases.
The delay of high energy photons relative to low energy ones in the X-ray variations
has also been revealed by the correlation analysis.
The implication of observed temporal and 
spectral characteristics to the production region and mechanism of Cyg X-1 X-ray 
variations is discussed.
\end{abstract}


\section{Introduction}

  Complex rapid fluctuation of the X-ray emission is a common characteristic  of 
Galactic black-hole candidates and low-mass X-ray binaries (\cite{klis95}). 
X-ray variabilities on different time scales carry valuable informations about X-ray 
emitting regions. Cygnus X-1 is the brightest source in the hard X-ray sky and the most
studied Galactic black hole candidate, its variability has been studied extensively.
The Fourier transformation is a powerful tool in timing analysis and a lot of  
important results on
Cyg X-1 variability, e.g., the shape of power spectral density, the coherence function
and time delays between soft and hard light curve variability, are derived with this
technique.  
Recently, with the Fourier analysis technique and observation data of the Proportional 
Counter Array (PCA) on board the {\sl Rossi X-ray Timing Explorer} ({\sl RXTE}),
a series of studies for different states of Cyg X-1 have been done, e.g.,
studies of temporal properties of Cyg X-1 in the soft state (\cite{cui97a}), 
during the spectral transitions (\cite{cui97b}), and in the hard state (\cite{now98}).        
On the other hand, studying temporal properties of Cyg X-1 has been done with 
analysis techniques in the time domain, e.g., studying the shot-noise model by 
the autocorrelation function (ACF) of the light curves (\cite{wei75,sut78,wei78,pri79}),
analysis of the cross-correlation function (CCF) between different energy bands
(\cite{pri79,nol81}), modeling the light curve of Cyg X-1 by an autoregressive process
(\cite{pot98}). The frequency in a Fourier power spectrum or cross spectrum does not
simply correspond to a certain time scale in the time domain which the physical process
causing the variation really occurs in.  
Timing analyzing directly in the time domain is
eventually necessary to study the variations on different time scales. 
  
  Temporal analyzing in the time domain usually requires the total amount 
of observed events and the signal to noise ratio of data much higher than that 
with the Fourier analysis. To overcome this difficulty, Negoro, Miyamoto and Kitamoto 
(1994) obtained average X-ray shot profiles from a {\sl Ginga} observation of 
Cyg X-1 in the low state by superposing many shots through aligning their peaks    
and studied their properties. Recently we made analysis of the structures of
 X-ray shots with the characteristic time scale $\sim 0.01-0.1$ s in the different states 
of Cyg X-1 with PCA/{\sl RXTE} observations 
and an improved searching and superposing algorithm (\cite{fe98}). 
From the obtained average shot profiles we have observed some interesting features, e.g.,
the average hardness of shot is lower than that of steady emission in the transition 
and hard states but higher than that in the soft state, the shot profile in the high 
energy band is narrower than that in the low energy band, etc. These observed phenomena 
give valuable hints on the shot production mechanism. 
In this work we study the 
correlation properties between the hardness and intensity and the energy dependence of 
temporal profiles on different time scales and for general variabilities, not only
for shots in $0.01-0.1$ s region.    

As chaotic fluctuations of X-rays from Cyg X-1 are exhibited on wide time scales, from hours 
down to smaller than 1 ms, it is difficult to study the variation phenomena on
a given time scale by simply using traditional analysis tools in the time domain,
e.g., the correlation coefficient, autocorrelation function  and 
cross-correlation function.
Large time bin used in calculation will erase the information on shorter time 
scales. And the analysis result with a short time bin reflects not only 
the variation property on the short time scale, but is also affected by that 
on longer ones up to the total time period used in the calculation.  
In this work we make timing analysis of PCA/{\sl RXTE} observations of Cyg X-1
by a modified approach in the time domain. For a given time bin $\Delta t$,  
we divide the total observation period into segments, each has a duration just 
several times of $\Delta t$, say 10$\Delta t$. The correlation coefficient,
ACF and CCF are calculated for each segment
separately and the results are finally averaged. The resultant correlation coefficient,
ACF and CCF, should 
be then related to the limited time scale region of $\Delta t - 10\Delta t$. 
In section 2 we present our analysis of correlation between hardness and intensity 
by this approach. The difference of ACF widths or time lag between low and high energy 
bands are usually much smaller than the time scale studied, which is another difficulty
for timing analysis in the time domain. We try to use a technique similar to that in
the vernier scale to measure the small difference in terms of a large time bin
and present the results on energy dependence of temporal profiles of Cyg X-1 X-ray 
emission on different time scales in section 3.      
The implications of observed temporal and spectral characteristics to the production 
region and mechanism of Cyg X-1 X-ray variations are discussed in section 4.
 
\section{Correlation between Hardness and Intensity}
A series of observations of Cyg X-1 had been performed by  {\sl 
RXTE} in 1996. On 1996 May 10 (131 day of 1996) 
the All Sky Monitor  
on {\sl RXTE} revealed the Cyg X-1 started a transition from the normal hard (low)
state to the soft (high) state. After reaching the soft state, it stayed in this state 
for nearly 2 months before going back down to the hard state (\cite{cui97b}). 
We use data of  
PCA/{\sl RXTE} (2--60 keV)   in 11 observational periods, including three periods 
during the hard-to-soft transition, three  the soft state, two  the
soft-to-hard transition and three  the hard state.  
The typical time duration of a period is, say, $\sim 2000$ s.
Two modes of PCA data are used in our analysis. One is the Event Mode with 16 energy 
bands covering 13.1--60 keV with the time resolution of 16 $\mu$s. The other is 
Single-Bit Mode including 2--6.5 keV and 6.5--13.1 keV bands (or 2--5 keV and 5--13.1 keV 
bands for the used data of hard states) with the time resolution of 125 $\mu$s.
The version 4.1 of standard {\sl RXTE} ftools is used to extract the PCA data
and version 1.5a of the background estimator program to estimate the background. 
The corrections of collimator response and barycentric time 
are applied to the chosen data of an observation period. The background-subtracted 
light curves of the low energy band $f_1$ (2--6.5keV) (or 2--5 keV for the hard state), 
middle energy band $f_2$ (6.5--13.1keV) and high energy band $f_3$ (13.1--60keV) 
with a selected time bin $\Delta t$ can be obtained by rebinning and making background correction for 
each energy band respectively.   
From the three light curves we derive the profiles of intensity $f$(2--60keV)
 and hardness ratio $h = f_3/f_1$ and   
use the correlation analysis to study the relationship between the 
 X-ray intensity and hardness ratio. The correlation coefficient 
for $n$ pairs of $h,f$ values is defined as 
\begin{equation}
r(h,f)=\sum_{i=1}^{n}(h(i)-\overline{h})(f(i)-\overline{f})/
\sqrt{\sum_{i=1}^{n}(h(i)-\overline{h})^{2}\sum_{i=1}^{n}(f(i)-\overline{f})^{2}}
\end{equation} 
   
For a given time bin $\Delta t$ the studied observation period is divided into 
total $N$ segments
with a duration of 10$\Delta t$ each. In our analysis, a data bin is defined as an 
effective one if the total length of data gaps in the bin is less than 0.1$\Delta t$,
and just such segments are used in which the number of effective bins $n \ge 8$.
We calculate the correlation coefficients $r(i)$ of each segment $i, ~i=1,2,..,N,$ 
by using Eq.(1), and then  
their average $\overline{r}=\sum_{i=1}^N r(i) / N$ and standard deviation 
$\sigma(\overline{r})=\sqrt{\sum_{i=1}^N (r(i)-\overline{r})^2/N(N-1)}$ 
for the period. 
Table 1, 2 and Figure 1 show the obtained averages of correlation
coefficients between $h$ and $f$(2--60keV) and their standard deviations for 
the 11 periods with time bin (time scale) 1 ms, 10 ms, 0.1 s, 1 s, 10 s, 20 s 
and 50 s  respectively, where the hardness ration $h$ is determined by the two
energy bands of 2--6.5 keV and 13.1--60 keV  for the periods of soft and transition states
and 2--5 keV and 13.1--60 keV for the hard states. The lower band range of Single-Bit mode 
of PCA observations for the hard state of Cyg X-1 in 1996 is a litter bit different
with that of public observations for the soft and transition states, 
this difference will not considerably change the results in this work.
 
Figure 2 (a)--(d) show the light curves and hardness ratio profiles in 50 s time bin
and their trends smoothed by the least-square fitting for four example periods
in different states.  
 Comparing the light curves (fluctuated solid curves) and the 
corresponding hardness ratio profiles (fluctuated dotted curves) in the four periods 
in different states, one can see that a near perfect correlation between the hardness
and intensity is shown in the upper right-hand panel of Fig.2 for the soft state,
 but rather complex feature,
a mixture of positive and negative correlations, for other states.   

The plots of correlation coefficient versus time scale for different states 
are shown in Figure 3. The value of correlation coefficient $r(h,f)$ in Fig.3 
on a time scale and for a given state is the weighted average  of 
$\overline{r}(k)$
from $k=1,..,M$ considered periods in this state, i.e., 
$<\overline{r}>=\sum_{k=1}^M \omega(k)\overline{r}(k)/\sum_{k=1}^M\omega(k)$, 
$\omega(k)=1/\sigma^2(\overline{r}(k))$ ; its error is determined by both the errors
of each $\overline{r}$ values and their dispersion, i.e. 
$\sigma=\sqrt{\sigma_1^2+\sigma_2^2},~ \sigma_1^2=1/\sum_{k=1}^M\omega(k),
~\sigma_2^2=\sum_{k=1}^M(\overline{r}(k)-<\overline{r}>)^2/(M-1)$.

 Some interesting features 
of state dependence of correlation coefficient $r(h,f)$ can be seen from Tables 1, 2 
and Figures 1--3. On the shortest time 
scale ($\Delta t=1$ ms) anti-correlations are always found in all observation periods. 
On the longer time scales (0.01--50 s)
the correlation property when Cyg X-1 in soft states is quite different
with what in transition and hard states. For soft states,
positive correlations between the hardness ratio $h$ and intensity $f$ exist 
on all time scales between 0.01 s and 50 s. The correlation
coefficients $r(h,f)$ increase monotonically and reach to near unity along with the time scale 
increasing from 0.01 s to 50 s. On the other hand, the correlation coefficients are 
negative or near zero when Cyg X-1 is in transition or hard states.
 The error bars in Tables 1, 2 and Figures 1, 3  are derived from the standard deviations 
of the obtained correlation coefficients  of all available segments.
 As the total number of correlation coefficients  is always large for any 
state and time scale studied, their deviation can give a reasonable estimate of
the statistical uncertainty in the derived average value.     
The increasing trend of the correlation coefficients with the time scale in the soft 
state and the difference between the correlation in the soft state and that in the other 
states are statistically significant.

\section{Energy Dependence of Temporal Profiles}
The Fourier frequency-dependent time delays of high-energy photons relative to low-energy 
ones in the X-ray variations of Cyg X-1 have been observed for some time
and are attracting an increasing amount of attention to using it as a diagnostic of the  
emission mechanism and emitting region (e.g., \cite{kaz97,hua97,bo98b,now99}).
Recently, from analyzing many X-ray shots of Cyg X-1 in different states we confirm the existence 
of time delays of hard photons behind soft photons in a shot (\cite{fe98}). 
A notable finding from our shot analysis is that the shot profile in the high energy 
band is narrower than that in the low energy band, which is not consistent with the 
prediction of the simple Comptonization models. It is important to study the energy dependence 
of temporal profiles of Cyg X-1 in different states and on different time scales directly 
in the time domain. 

We take a procedure similar to that used in section 2 to derive the average width 
of the autocorrelation functions
and study the temporal correlation of two light curves in different energy bands
on different time scales. For a given time scale $\Delta t$, the observation period 
studied is also divided into segments with a duration of $10\Delta t$ each.
The autocorrelation  function of the zero-mean time series 
${v(i)=f(i)-\overline{f}}$ of a segment is usually defined as
\begin{equation} ACF(k) = \sum_i v(i)v(i+k)/\sigma^2(v) \hspace{5mm} (k=0, \pm1, ...) 
\end{equation}
where $\sigma^2(v)=\sum_i (v(i))^{2}$, $f(i)$ is the average intensity in the time bin 
$(i-1)\Delta t - i\Delta t$. 
The cross-correlation function of two series $v_{l}$ and $v_{m}$ corresponding 
to two energy bands is usually defined as
\begin{equation} CCF_{l,m}(k) = \sum_i v_{l}(i)v_{m}(i+k)/\sigma(v_l)\sigma(v_m) 
\hspace{5mm} (k=0, \pm1, ...) \end{equation}
The corresponding time lag of $ACF(k)$ or $CCF(k)$ is $\tau=k\Delta t$. With the $ACF$ and $CCF$
defined above, it is difficult to measure time lags $\tau \leq \Delta t$. 
To get necessary resolution for time lags we modify the above definition of $ACF$ and $CCF$
that we can use a  time bin $\delta t < \Delta t$ for time lag.
For the convenience we  use the time instead of the bin number
as the argument  in the expression of $ACF$ and $CCF$: use $ACF(\tau)$ and $CCF(\tau)$ 
to represent the values of $ACF$ and $CCF$  at the time lag $\tau$, $f(t)$ 
the average intensity in
the time duration $t-t+\Delta t$ and $v(t)=f(t)-\overline{f}$. We define 
the autocorrelation function of the light-curve $f$ 
\begin{equation} ACF(\tau)=\sum_{i}v(i\Delta t)v(i\Delta t+\tau)/\sigma^2(v) \end{equation}
and the cross-correlation function between two light-curves $f_l$ and $f_m$
\begin{equation}
 CCF_{l,m}(\tau)=\sum_{i}v_l(i\Delta t)v_m(i\Delta t+\tau)/\sigma(v_l)\sigma(v_m) 
\end{equation}  

With the $ACF$ and $CCF$ at $\tau=k\delta t, k=0, \pm1, ...$, we can study temporal variations 
on the time scale
of $\Delta t$ -- $10\Delta t$ with a necessary time resolution $\delta t$.  
For a studied observation  and a given time bin $\Delta t$, the total observation
period is divided into $N$ segments with a duration of $10\Delta t$ each. 
For a segment $j$  we calculate the autocorrelation function   
of the light curve in the low energy band, $ACF_1$,  and that in the high energy band,
$ACF_3$, with the time lag bin $\delta t$ of 1 ms from Eq.(4),
and also $CCF_{1,3}$ between the two bands from Eg.(5).
We then find out the full width at half maximum, $W_{1}(j)$, from the
profile $ACF_1$, $W_{3}(j)$ from $ACF_3$, and the time delay $\lambda(j)$ 
of the high energy band relative to the low energy band.  The time delay is 
determined by the maximum of 
the function $CCF_{1,3}(\tau)/CCF_{1,3}(\tau=0)$.
The $N$ segments are divided into $M=10$ groups with the segment numbers $n(k)$ in each group 
$k ~(k=1,2,...,M)$ being nearly equal to each other and $\sum_{k=1}^{M}n(k)=N$. 
The average of a concerned quantity $x$ ($x$ can be $W_1, W_3$ or $\lambda$) and 
its standard deviation for each group $k$ can be calculated by using the following formulas
\begin{equation}
\overline{x}(k)=\sum_{j=1}^{n(k)}x(j)/n(k), \hspace{7mm}
 \sigma(\overline{x}(k))=\sqrt{\frac{\sum_{j=1}^{n(k)}(x(j)-\overline{x}(k))^2}
{n(k)(n(k)-1)}}
\end{equation}
Finally we report the result for the observation as
\begin{equation} x = <x> \pm \sigma_1(x) \pm \sigma_2(x)  \end{equation}
where
\[ <x> = \sum_{k=1}^{M}\omega(k) \overline{x}(k) / \sum_{k=1}^{M}\omega(k), \hspace{7mm}
\omega(k)=1/\sigma^2(\overline{x}(k)) \] 
\begin{equation} \sigma_1(x)=\sqrt{1/\sum_{k=1}^{M}\omega(k)}~, \hspace{7mm} 
\sigma_2(x)=\sqrt{\sum_{k=1}^{M}(\overline{x}(k)-<x>)^2/M(M-1)} \end{equation}
The first error term $\sigma_1(x)$ in the expression (7) represents the precision 
of $<x>$ as an average for the observation period, the second one $\sigma_2(x)$ 
describes the dispersion of the $M$ averages $\overline{x}(k)$ within the observation.

 We apply the formulas (6) -- (8) to calculate  the average $ACF$ widths
$FWHM_1 = <W_1>$ in the low energy band, $FWHM_3 = <W_3>$ in the high energy band, and 
the average time delay $<\lambda>$ between the two bands for six observations
and show the results in Table 3 -- 5. From Table 4 and 5 one can see that 
for all studied observations  the temporal variations
in the high energy band are always delayed relative to the low energy band
on all time scales, and that 
 on the time scale $\Delta t=0.1$ s the width of $ACF$ in the high energy band 
is generally narrower  than that in the low energy band.
For an example, Figure 4 shows  the average autocorrelation and cross-correlation 
functions of the low and high energy light-curves with the time bin $\Delta t=0.1$ s 
for the observation started at 143.7 
day of 1996 when Cyg X-1 in the transition from hard state to soft state. 
The total duration of the observation is about 2230 s. The total effective 
bins of 0.1 s are divided into $M=10$ groups, each group has about $n=208$ segments  
of 1 s.  For the 10 groups, the $ACF$ width ratios, $W_3/W_1$, are all smaller
than unity , they are $0.797\pm0.029,
0.755\pm0.030, 0.746\pm0.031, 0.775\pm0.033, 0.725\pm0.030, 0.715\pm0.029,
0.752\pm0.027, 0.749\pm0.030, 0.739\pm0.028$ and $0.712\pm0.030$.
From the above values we finally obtain $FWHM_3/FWHM_1=0.747\pm0.009\pm0.016$.
The above results show that the width of $ACF$ in the high energy 
band being narrower than that in the low energy band is statistically significant
for the observation.
To compare our results on hard X-ray delay with that from Fourier spectral 
analysis, we plot the time lags as a function of $1/\Delta t$ for four
observations in Figure 5 (a)--(d). The time lags vs. $1/\Delta t$ shown in Fig.5
are similar with the time lags vs. Fourier frequency obtained by 
Fourier spectral analysis (\cite{cui97b}, Fig.8; \cite{now98}, Fig.10).

\section{Discussion}  
There exists an obvious anti-correlation between the 2-60 keV  intensity  
and  spectral hardness of  X-rays from Cyg X-1 in  its long-term variability    
exhibited by its state transition between low/hard and high/soft states.
One can find from Fig.2 that for the soft state the typical intensity  
is $\sim 15$ cts/ms and hardness ratio between the high and low 
energy bands $\leq 0.1$, and for the hard state they are $\sim 6$ and $\geq 0.6$,
respectively. For shorter time scales, our analysis results show different 
correlation features in different states.

In a previous paper (\cite{fe98}) we studied the spectral properties of average 
shots of Cyg X-1 on the time scale of about 0.1 s 
in soft, hard and transition states and found that the hardness during a shot is 
higher than that of the steady component around the shot in soft states but lower
in transition and hard states. 
Fig.6(a) -- Fig.6(d) show the intensity and hardness ratio profiles of average shots 
in the different states of Cyg X-1, where
the average shots were obtained by using the same data as that presented in this paper.
We can see from Fig.6(b) that for the soft state the hardness ratio peaks while 
the flux of the shot peaks, showing a positive correlation between 
the hardness and intensity. For the hard state, although the average hardness of 
the shot is lower than that of  the steady emission, but there exist both positive 
and negative peaks during the shot, as shown in Fig.6(d), the net correlation 
for the shot should be quite small.  For both the hard to soft transition state 
and the soft to hard transition state, the hardness variation during the shot 
is dominated by a negative peak, as shown in Fig.6(a) and Fig.6(c), 
the correlation coefficient should be negative.
In fact we can calculate the correlation coefficient between the profiles of flux 
and hardness ratio of a shot. The derived $r(h,f)$ for 
shots in Fig.6 within the range of $\pm 0.2$ s around the shot peak  are
$-0.44\pm 0.14$ for the hard to soft transition state, $0.89\pm 0.07$ for the 
soft state, $-0.38\pm 0.15$ for the soft to hard transition state, 
and $-0.04\pm 0.16$ for the hard state, respectively. The state dependence of   
shot $r(h,f)$ is generally consistent with what is shown in Table 1 and Fig.3 
for the time scale of 0.1 s, except a considerable difference in quantity    
for the soft state. The shot of the soft state shows a nearly perfect correlation 
between the hardness ratio and flux, but the correlation coefficients for the soft state
on 0.1 s time scale in Table 1 and Fig.3 are just about 0.3.  As the correlation 
analysis in this paper is made for the temporal variability on a certain time scale
in general, not only for shots, the obtained  correlation coefficient being smaller 
than that of shots reflects the existence of other uncorrelated variation
components, including noise. For the hard and transition states, the consistency 
between the results from our general correlation analysis on the 0.1 s time scale 
and those for shots indicates the variable flux on the 0.1 s time scale being dominated 
by shots. The trend of the correlation
coefficients $r(h,f)$ increasing monotonically and reaching to near unity along with 
the time scale from 0.01 s to 50 s may indicate the effect of uncorrelated
noise being weakened on larger time scales.

The most remarkable result from our analysis is the state dependence of correlation features
on the time scales of $\sim 0.01$ s$ - \sim 50$ s. 
The correlation coefficients $r(h, f)$ are always positive  
during soft states but negative or near zero during transition and hard states.
 It is widely believed that hard X-rays of Cyg X-1 come from inverse Compton scattering 
of low energy seed photons by high energy  electrons in an accretion disk corona (ADC) 
(e.g., \cite{ear75,pou96,do97b})
or hot advection-dominated accretion flow (ADAF) (e.g., \cite{es97}).  
The steady or slowly varied component of
the observed hard X-rays should be a global average of emission from different regions 
of the hot corona (ADC or ADAF), whereas a rapid variation can be produced at a local 
region where a disturbance occurs. 
The innermost region of the cold disk jointed with the hot 
corona is most probably a turbulent region where violent disturbances frequently 
take place and produce rapid varying low energy photons, causing variable X-ray
emission through Comptonization in the nearby region of the hot corona.
The difference between the hardness ratios of the rapid varied component and steady component   
may be caused by a difference between the local temperature of the Comptonizing region
to produce the rapid varied emission and the average temperature of the total corona.

 Two different geometries in the ADC or ADAF model exist. One is the sphere+disk 
geometry: a spherical corona around the centric compact object with exterior 
accretion disk. And another is the slab geometry: the Comptonizing medium being
a slab surrounding the disk. 
The predicted spectrum of Comptonization models with the sphere+disk 
geometry (\cite{do97b,es97}) can describe the observed X-ray spectrum 
of Cyg X-1 in the hard and transition states. The spherical corona exists only for
accretion rates below some critical value. For the soft state, the disk extends 
down to the last stable orbit and the hot flow is restricted to the corona 
surrounding the disk (\cite{es97}). 
The temperature $T_d$ of the disk is a function of radius: inner part 
has higher temperature (\cite{fra92,do97b}).
The temperature $T_c$ of corona embedding the cold disk is related to the 
disk temperature $T_d$:
for the same value of the optical depth of the corona, the spectra of Comptonized photons 
are harder in the case of higher $T_d$ (\cite{haa93,do97a}). 
Accordingly in the soft state with the slab geometry, the spectra of shots from the coronal region 
surrounding the innermost disk should be harder. On the other hand, 
for the transition and hard states the radial dependence of the coronal temperature 
in the central corona 
expected for an advection dominated accretion flow can be approximately represented  as 
$T_{c} \propto r^{-1}$ with $r$ being the distance from the center of the black hole
(\cite{nar94,nar98}). 
The innermost orbit of the disk in the case of the sphere+disk geometry is located outside 
the spherical corona and its neighboring Comptonizing region is the outer part of the corona
with a temperature lower than the average of the total corona. 
Thus, assume most of observed rapid variations 
take place 
at the joint region between the innermost disk and hot corona, the observed correlation
character will be consistent with what is expected by the models mentioned above, and, 
in consequence, the average correlation coefficient of time segments with limited length  
will be a useful parameter to probe the corona geometry of X-ray binaries.  

Another noticeable fact in our results is that the correlation coefficients $r(h,f)$
on the 1 ms time scale listed in the third column of Table 1 
are all negative, regardless what state Cyg X-1 was in.
It is most probably an indication of the effect of Compton cooling of the local
Comptonizing region by brief flashes. A considerable part 
of  rapid variability of Cyg X-1  consists of many aperiodically occurring shots with   
widths distributed between $\sim30$ ms and $\sim0.2$ s (\cite{fe98}).
We used time segments of 10 ms duration to determine the correlation coefficients
in the case of 1 ms time bin.
In case of a 10 ms segment being contained by a shot, the average hardness in our 
correlation analysis reflects not the  
average temperature of total corona  but that of local Comptonizing region of the shot.     
For this case a brief flashes of seed photons will decrease the local temperature 
and then the hardness of emitted x-rays.
                                                                                                                                     
 Temporal structures on the time scale of 0.1 s in the high energy band are  narrower 
than that in 
the low energy band as shown in Table 4, which is consistent with the average shot 
structures on the time scale $\sim0.1$ s in different energy bands obtained
from PCA/${\sl RXTE}$ observations of Cyg X-1 in different states (\cite{fe98}),
but contrary to the prediction of Comptonization models 
(\cite{su80,hu97,bo98a}).  This observed fact, together with the characteristic correlations
between hardness and intensity revealed by our correlation analysis and shot analysis
in the time domain, has to be considered in modeling the production process
of X-rays from Cyg X-1.                                                                                                                                                                                                                                                                                                                                                                                                                                                                                                                                                                                                                                                                                     
The hard X-ray time lag of Cyg X-1, expected by Comptonization process and revealed by 
 Fourier frequency analysis  is also shown in our results. 
It is interesting to see the  time lags as a function of $1/\Delta t$ (Fig.5)  
similar to that as a function of Fourier frequency (see, e.g., \cite{now98}, Fig.10).
In comparison with the Fourier frequency, the time scale $\Delta t$ is more directly 
related to the real physical processes causing the rapid variabilities.
At large time scales, $\Delta t \ge 1$ s, average lags $\overline{\lambda}$ change 
significantly with time, causing large $\sigma_2$ in the last column of Table 5
and large error bars in the region of $1/\Delta t \leq 1$ s$^{-1}$ in Figure 5.
One can see from Fig.5 that the correlation analysis technique in the time domain 
we proposed has the capability to study the time lags at the short time scale region 
(or "high frequency" region $1/\Delta t \ge 10$ s$^{-1}$) with an enough precision.
In this paper we use time lag bin $\delta t=1$ ms. A work of using higher time 
resolution data of ${\sl RXTE}$ to study time lags at the region
$1/\Delta t > 100$ s$^{-1}$ and correlations between hardness and intensity on time scales smaller than 1 ms   
is in the process, the results will be reported in a separate paper.
 
\acknowledgments 
The authors thank Dr. Yu Weinfei for helpful discussions. We also thank the referee
for valuable comments and suggestions. This research was supported by the National Natural
Science Foundation of China and  made use 
of data obtained through the High Energy Astrophysics Science Archive Research 
Center Online Service,
provided by the NASA/Goddard Space Flight Center.


\clearpage
\begin{figure}
\epsscale{1.0}
\plotfiddle{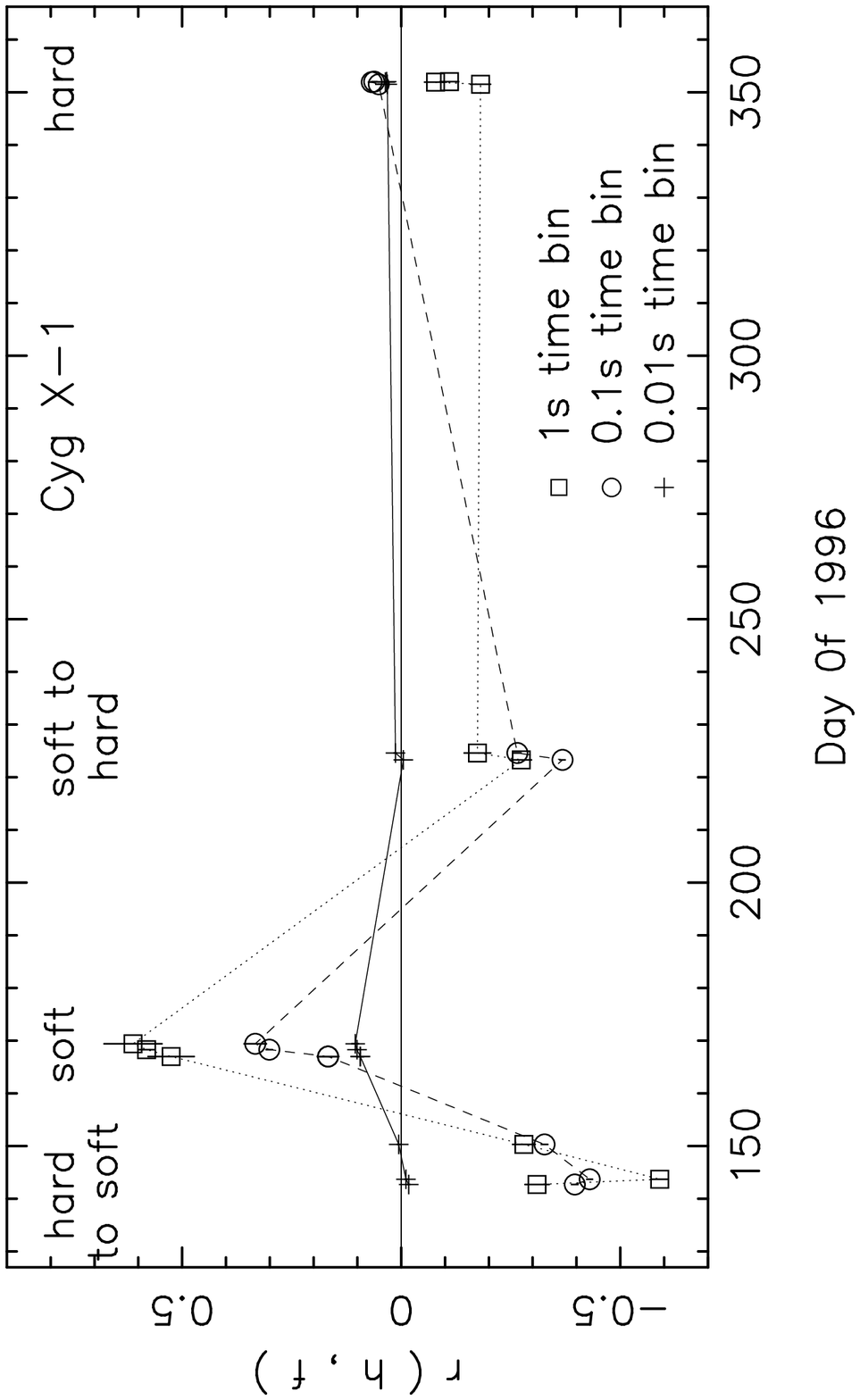}{20pt}{-90}{36}{60}{-300}{200}
\plotfiddle{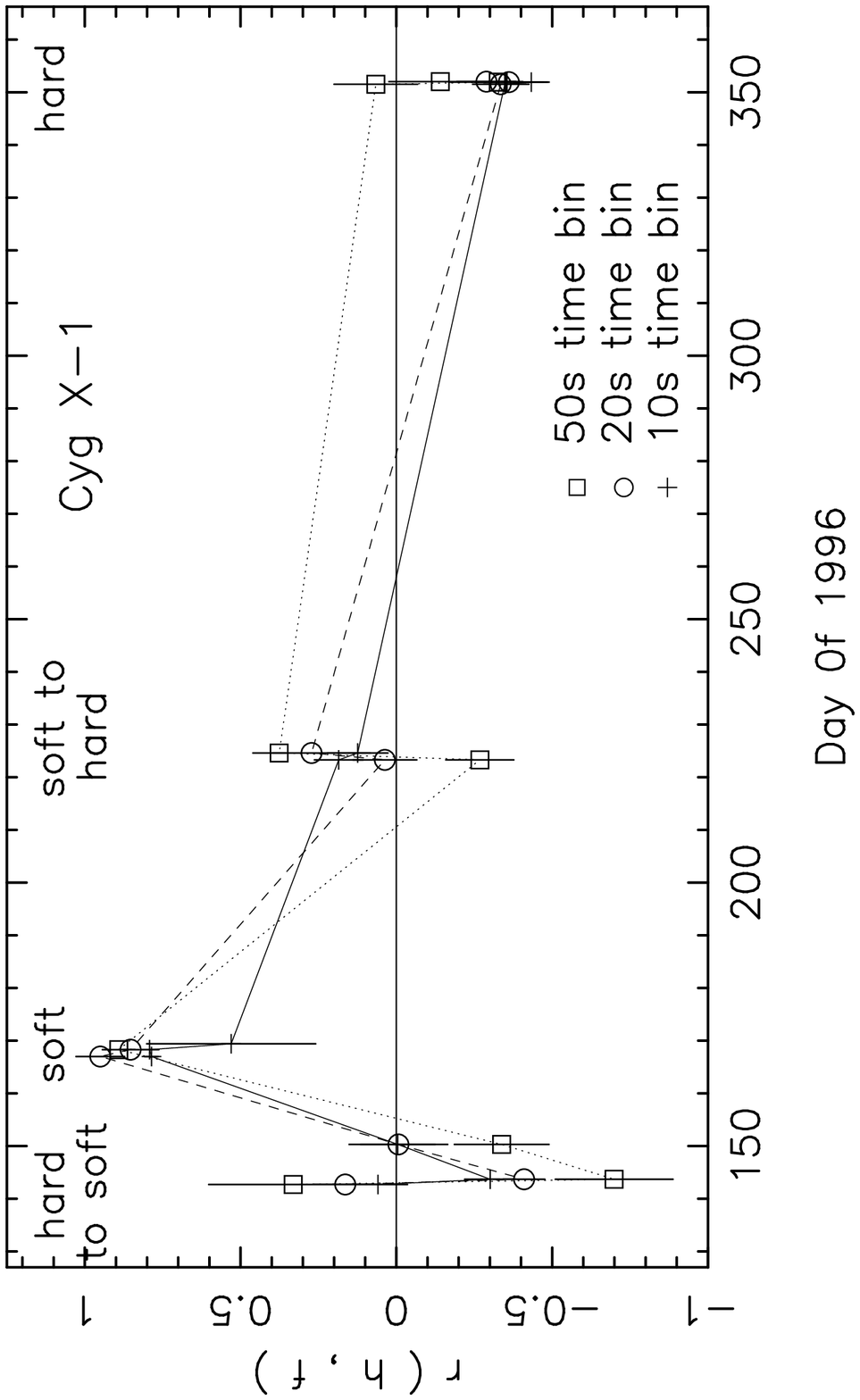}{20pt}{-90}{36}{60}{-50}{235} 
\vspace{4cm}\caption{ Correlation coefficients between hardness ratio $h$ 
and intensity $f$(2--60 keV). For the periods of soft and transition states,
$h=f$(13.1--60 keV)/$f$(2--6.5 keV). For the periods of hard states,
$h=f$(13.1--60 keV)/$f$(2--5 keV). (a) in time bins 0.01 s, 0.1 s and 1 s;
(b) in time bins 10 s, 20 s and 50 s. \label{fig1}}
\end{figure}

\clearpage
\begin{figure}
\epsscale{1.0}
\plotfiddle{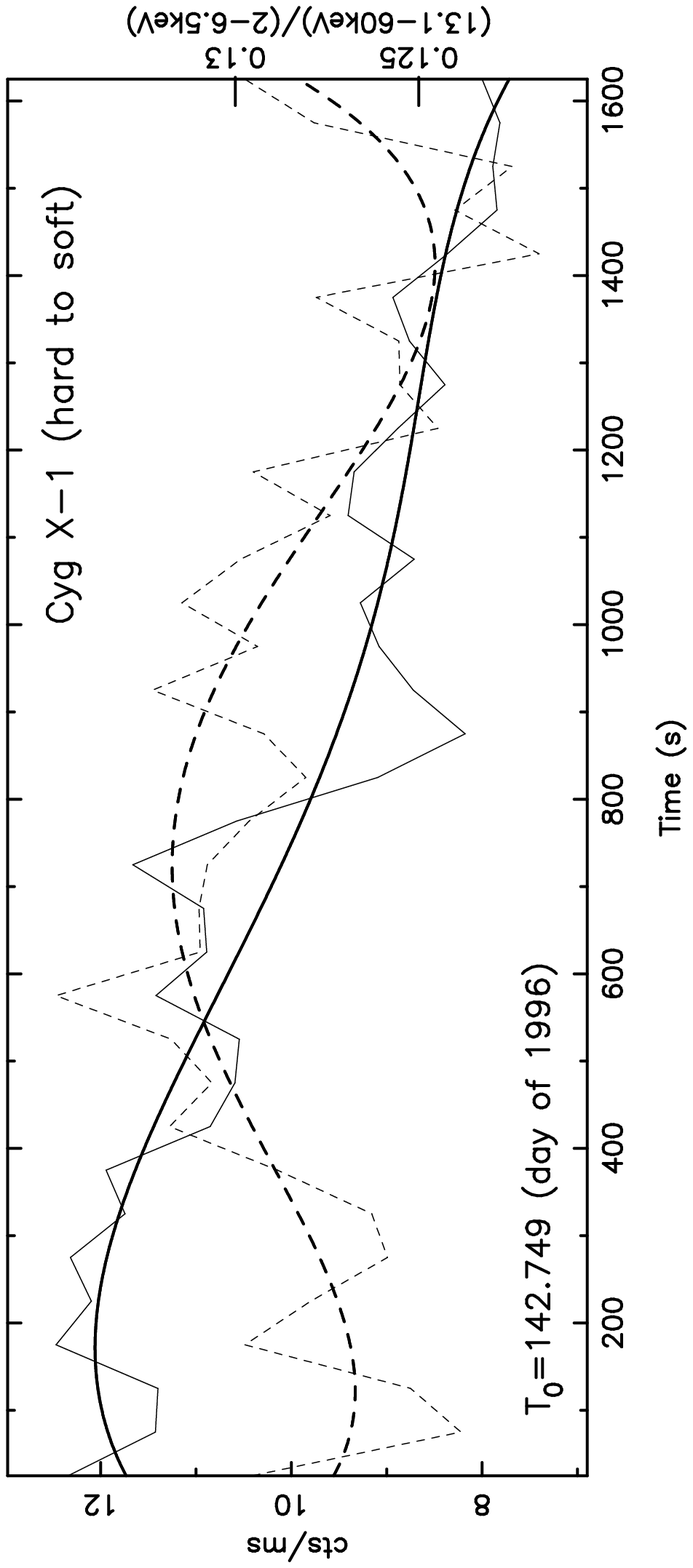}{20pt}{-90}{36}{50}{-300}{220}
\plotfiddle{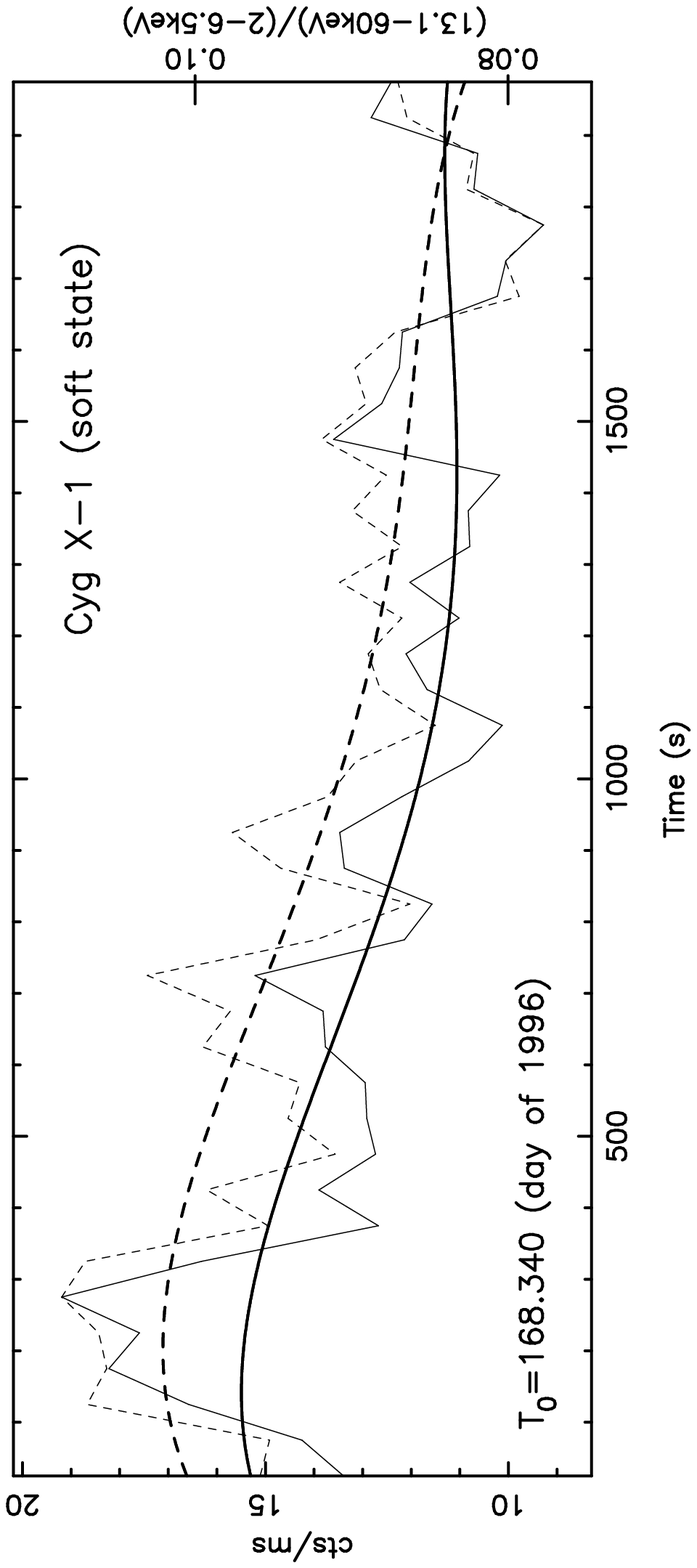}{20pt}{-90}{36}{50}{-50}{255}
\plotfiddle{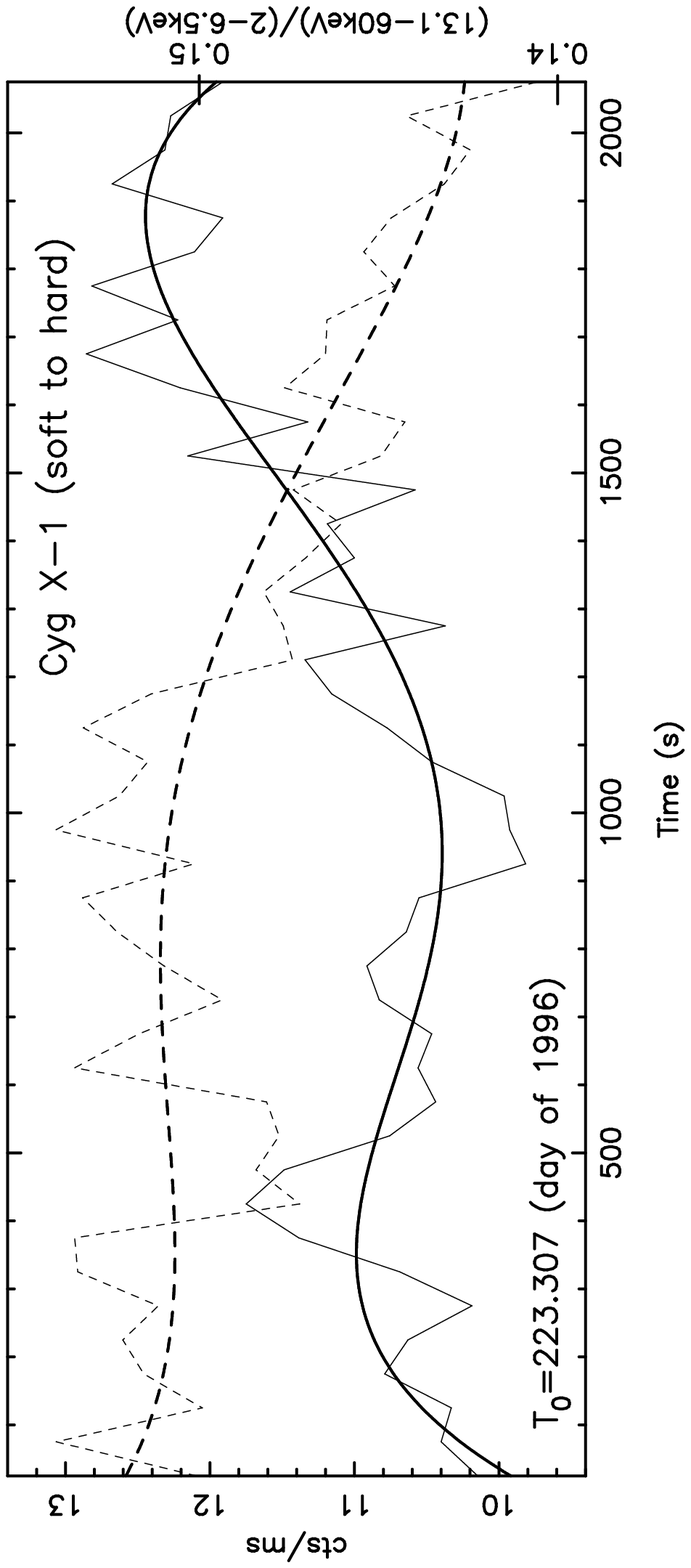}{20pt}{-90}{36}{50}{-300}{150}
\plotfiddle{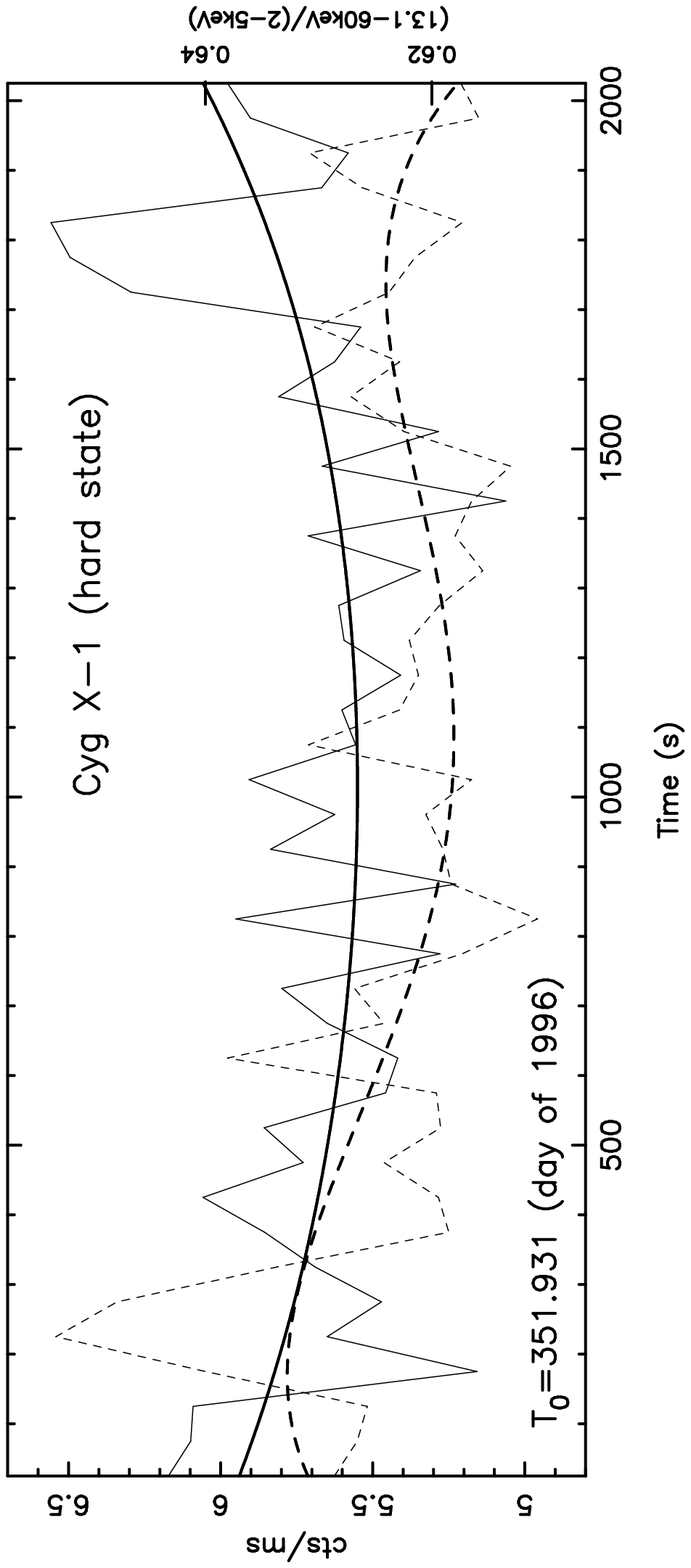}{20pt}{-90}{36}{50}{-50}{185}
\vspace{4cm}
\caption{Intensity and hardness ratio profiles 
of four observation periods
in different states. (a) Upper-left panel: hard-to-soft transition.
(b) Upper-right panel: soft state. (c) Lower-left panel: soft-to-hard transition.
(d) Lower-right panel: hard state. 
The solid fluctuated lines represent the light curve of 2--60 keV energy band in 50 s 
time bin, and the dotted fluctuated lines are the corresponding
hardness ratio profiles of 13.1--60 keV and 2--6.5 keV (or 2--5 keV for the period of
hard state). The solid and dotted smoothed lines are the least-squares fits of the
intensity and hardness ratio distributions respectively. \label{fig2}}
\end{figure}

\begin{figure}
\epsscale{1.0}
\plotfiddle{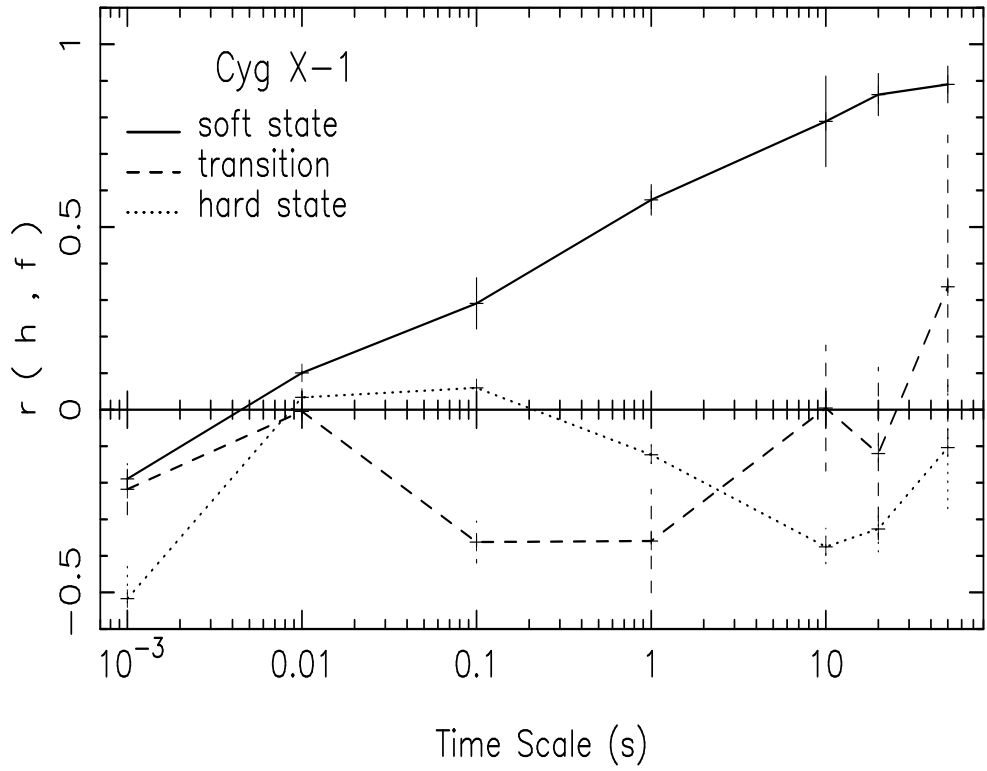}{20pt}{-90}{60}{80}{-300}{200}
\vspace{10cm}
\caption{Correlation coefficient between
hardness and intensity vs. time scale. Solid: soft state; Dashed: intermediate state;
Dotted: hard state. \label{fig3}}
\end{figure}

\begin{figure}
\epsscale{1.0}
\plotfiddle{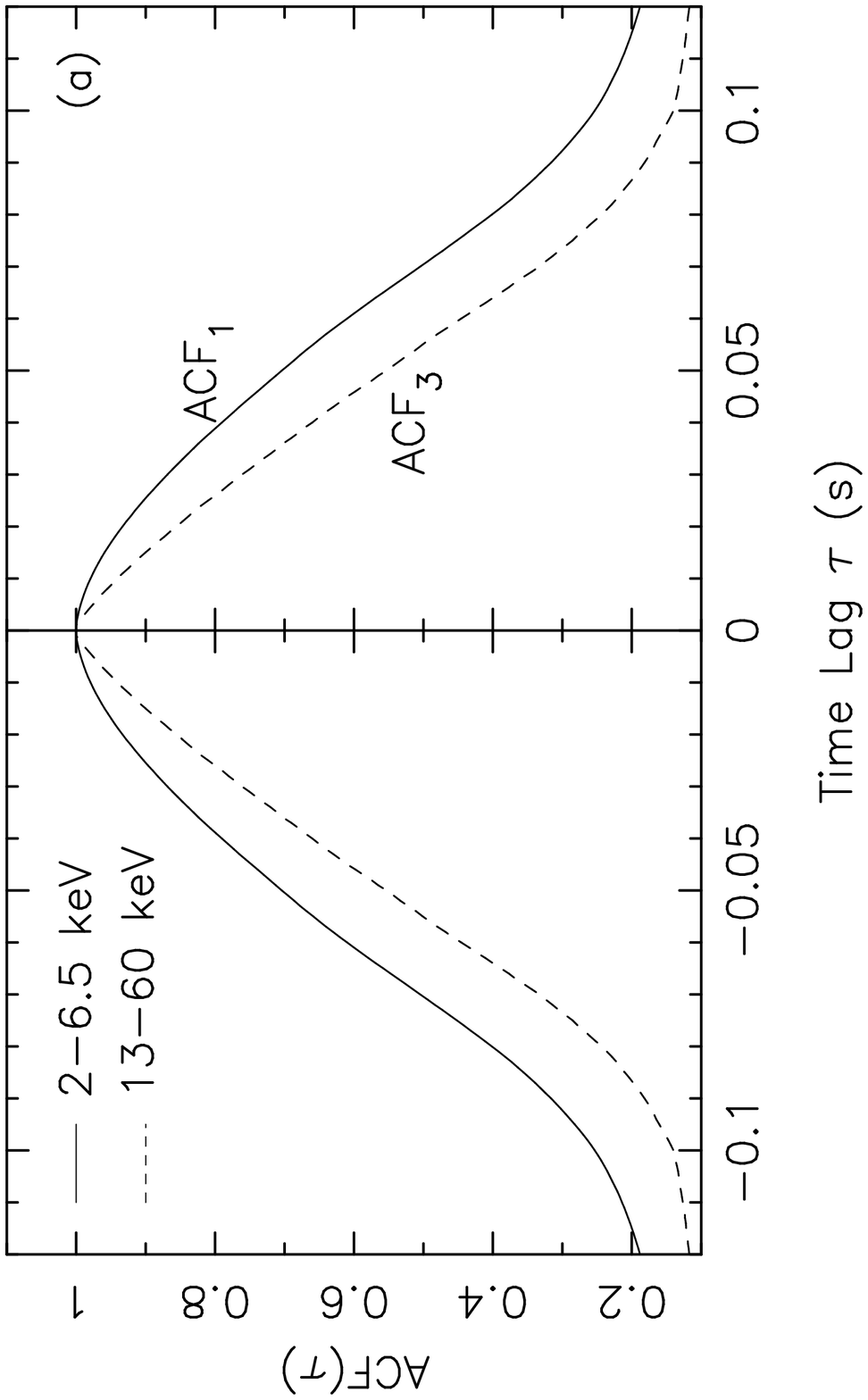}{20pt}{-90}{36}{60}{-300}{200}
\plotfiddle{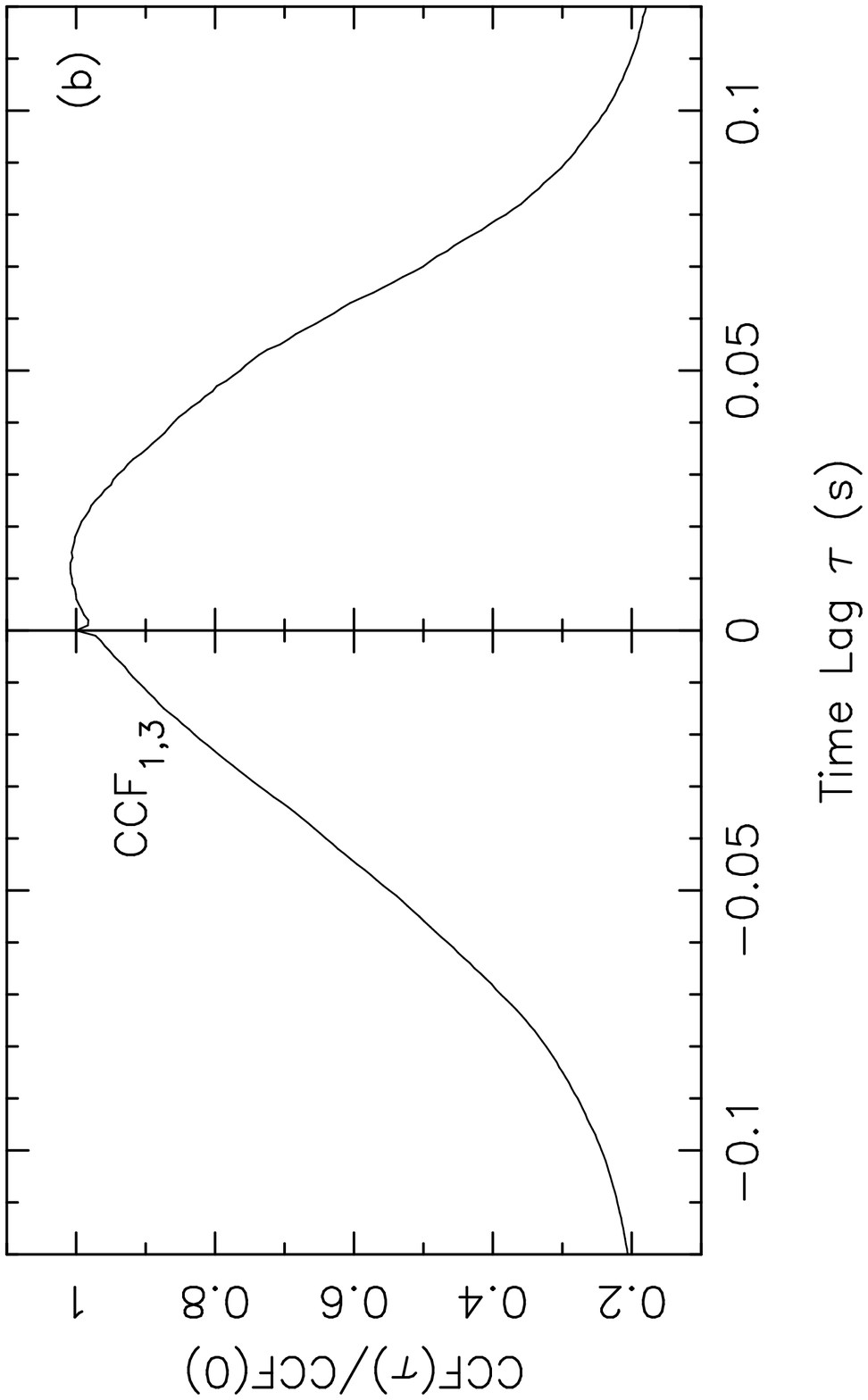}{20pt}{-90}{36}{60}{-50}{235}
\vspace{4cm}\caption{ The average auto-correlation and cross-correlation 
functions of an observation period of Cyg X-1 in the transition from the
hard state to the soft state.  The observation started at 143.7 day of 1996.
The time bin of light-curves $\Delta t=0.1$ s.
(a) Left panel: the solid line is $ACF$ of the low energy band (2--6.5 keV),
the dashed line is $ACF$ of the high energy band (13.1--60 keV).
(b) Right panel: $CCF$ between the low and high energy light-curves, normalized
at  the point $\tau=0$. \label{fig4}}
\end{figure}

\begin{figure}
\epsscale{1.0}
\vspace{5.5cm}
\plotfiddle{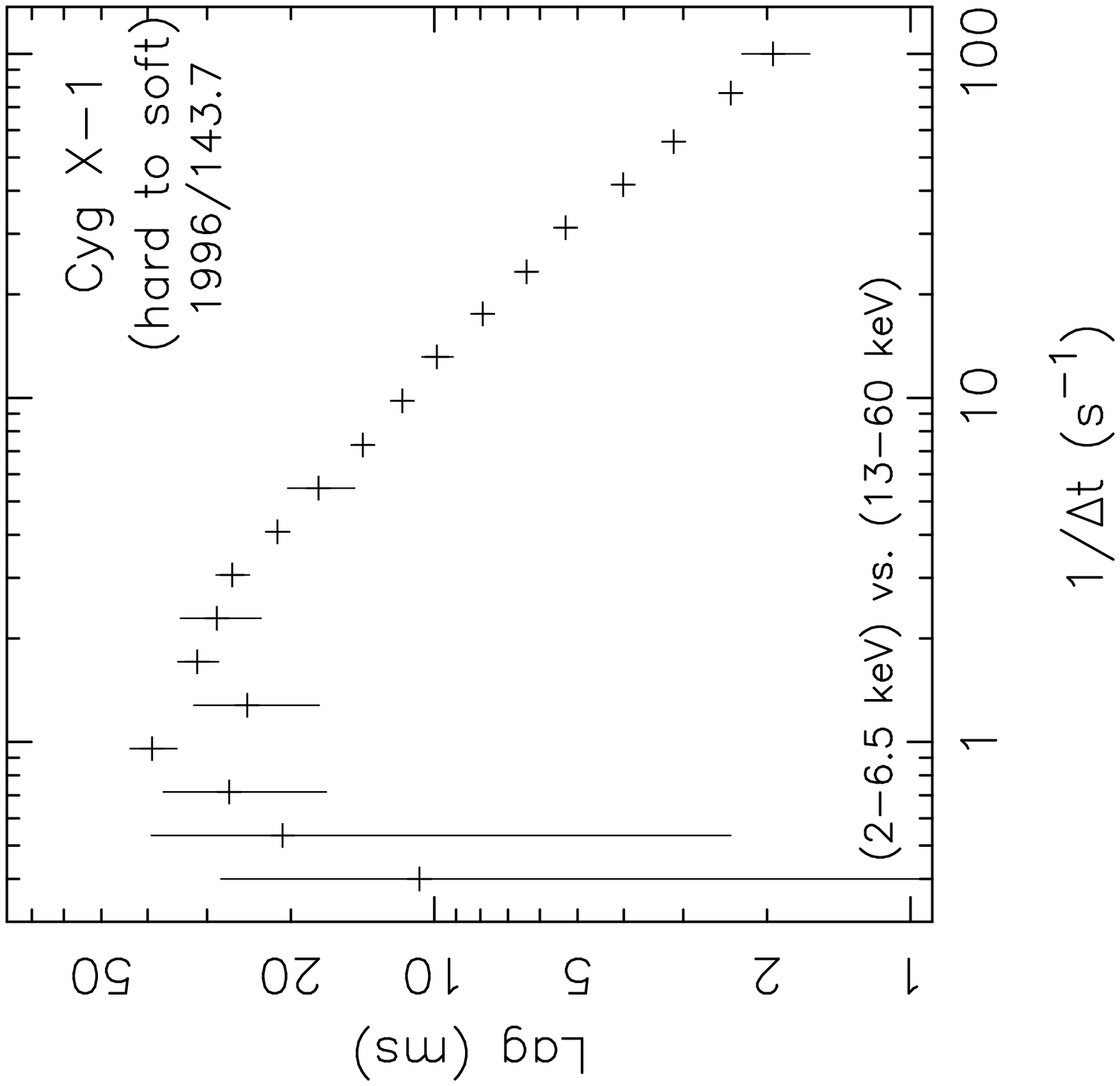}{20pt}{-90}{36}{30}{-250}{220}
\plotfiddle{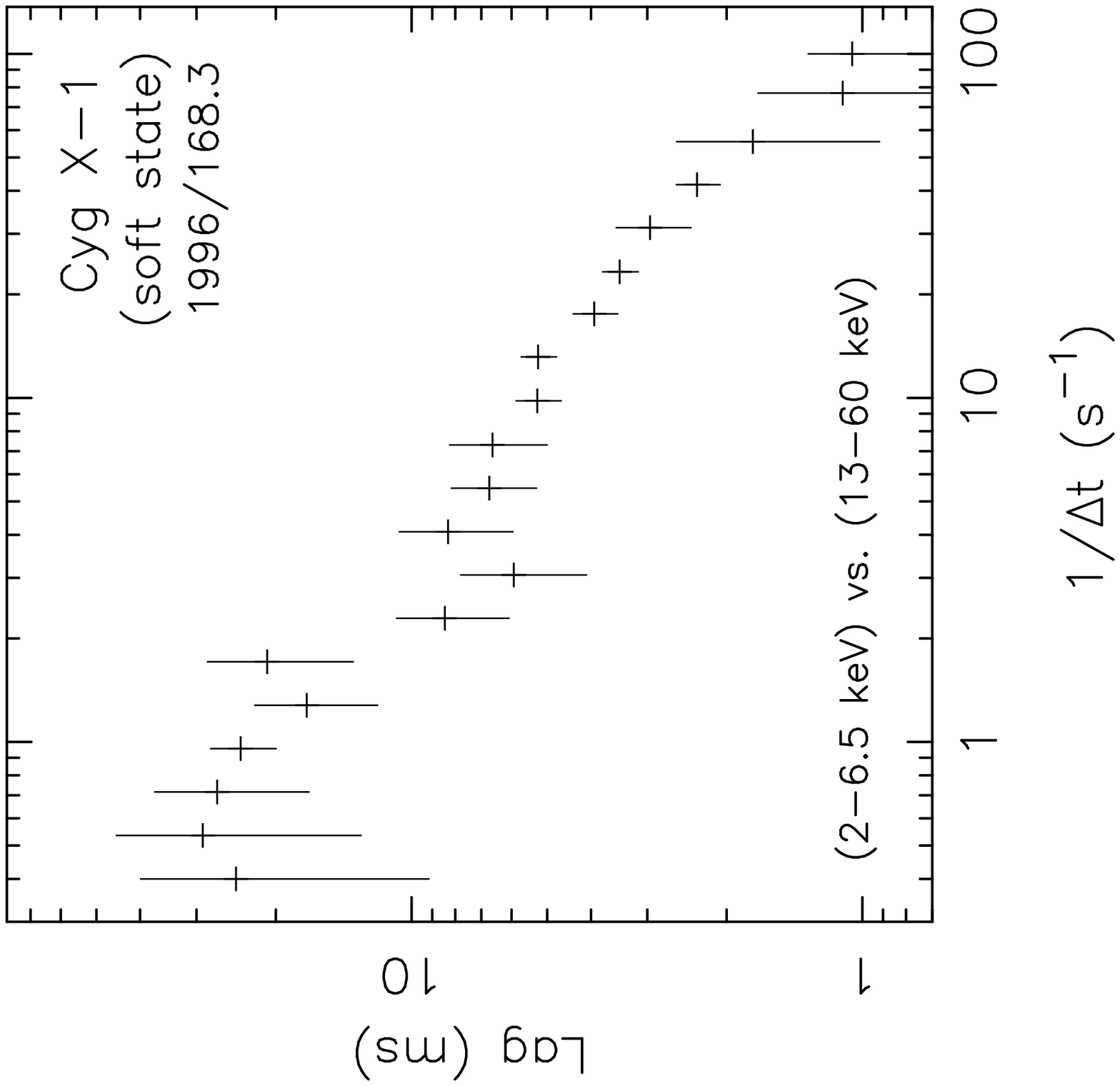}{20pt}{-90}{36}{30}{-50}{255}
\vspace{7mm}
\plotfiddle{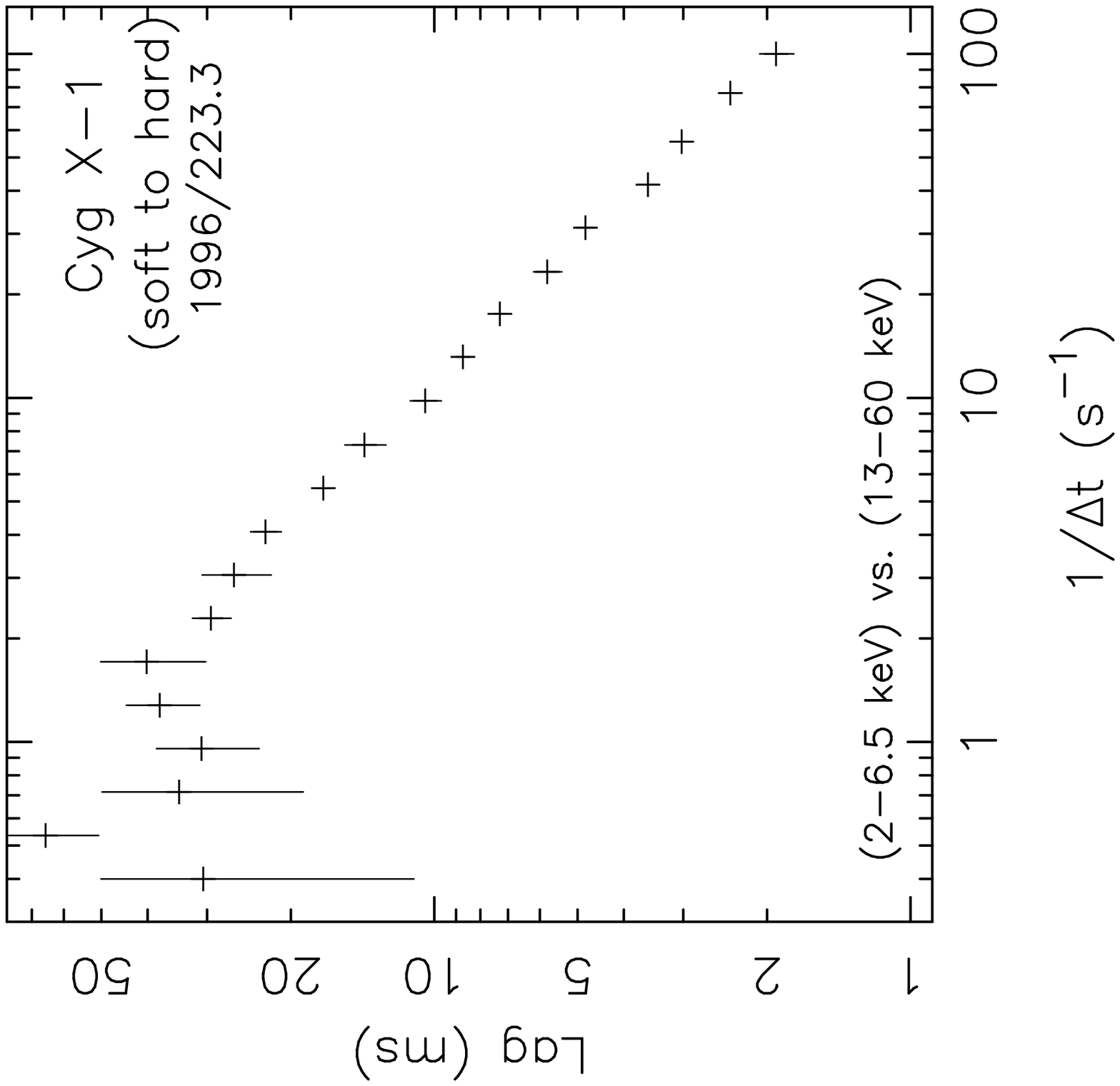}{20pt}{-90}{36}{30}{-250}{150}
\plotfiddle{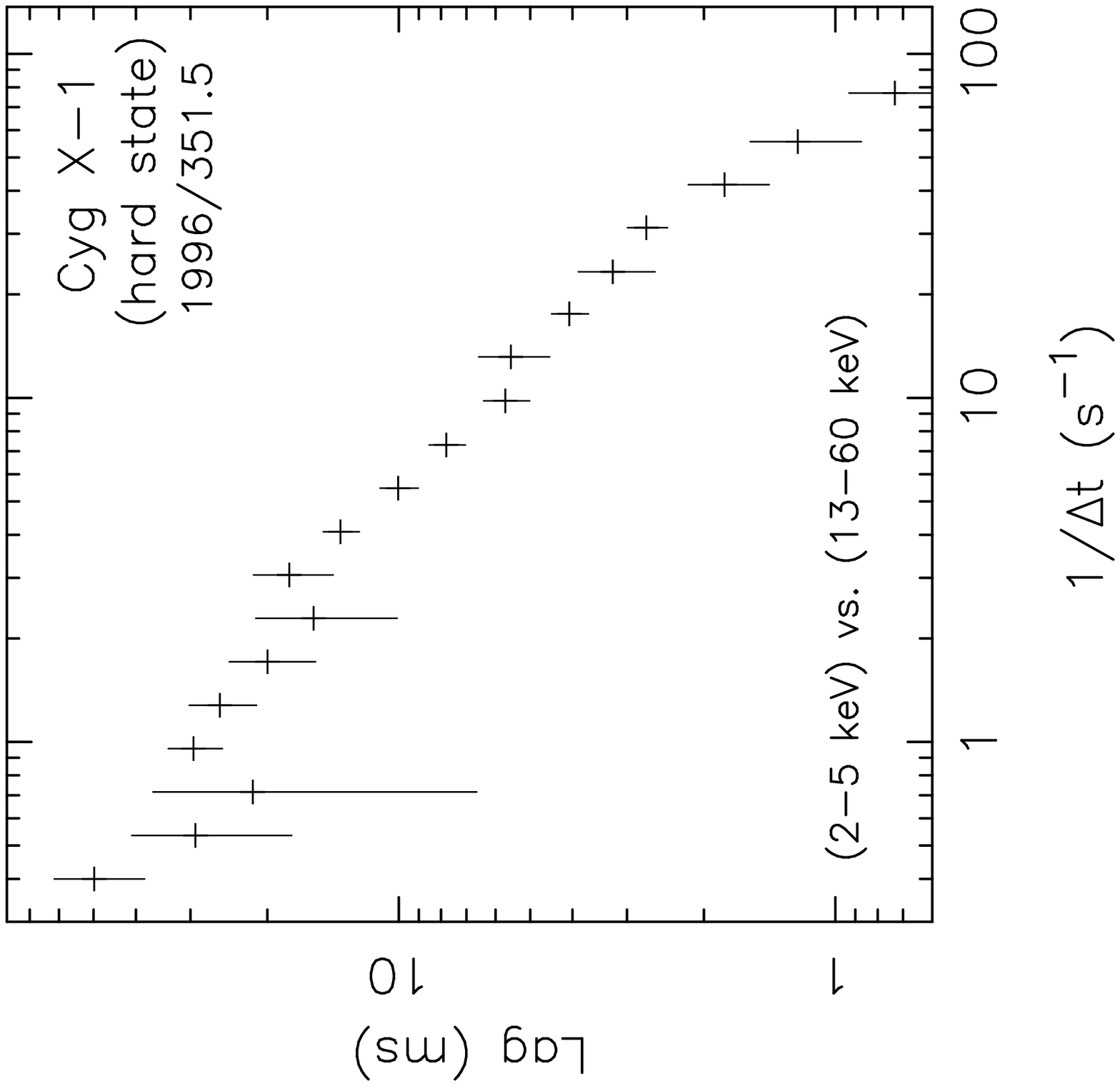}{20pt}{-90}{36}{30}{-50}{185}
\vspace{-5mm}
\caption{Hard X-ray lags between X-ray variations in the low energy and  high
energy bands of Cyg X-1 (2--5 keV vs. 13.1--60 keV for the hard state, 2--6.5 keV 
vs. 13.1--60 keV for the other states). The time lags are derived  by 
the cross-correlation technique in the time domain  for different time scales 
$\Delta t$. The error bars  $\sigma = \sqrt{\sigma_1^2 + \sigma_2^2}$,
where $\sigma_1$ and $\sigma_2$ are calculated by Eg.(8).
  (a) Upper-left panel: hard-to-soft transition.
(b) Upper-right panel: soft state. (c) Lower-left panel: soft-to-hard transition.
(d) Lower-right panel: hard state.  \label{fig5}}
\end{figure}

\newpage
 \begin{figure}
\epsscale{1.0}
\plotfiddle{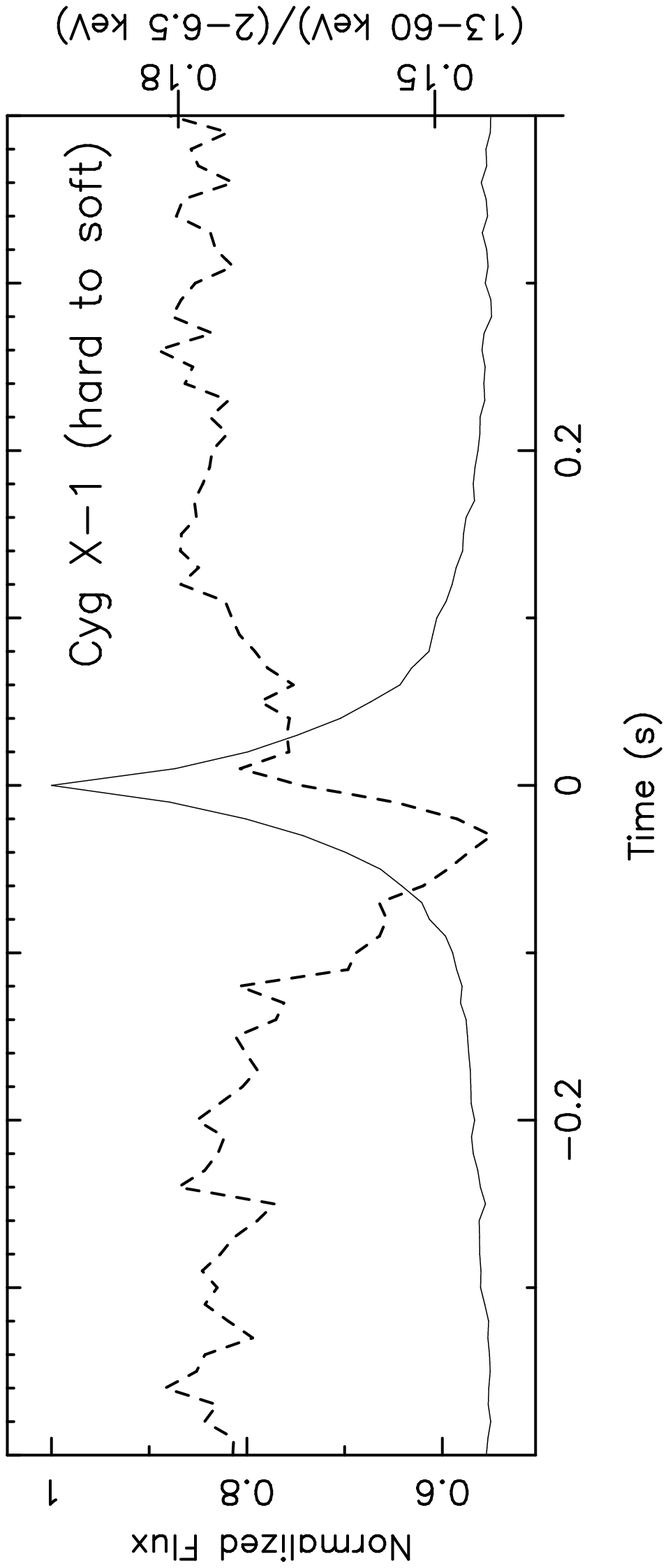}{20pt}{-90}{36}{50}{-300}{220}
\plotfiddle{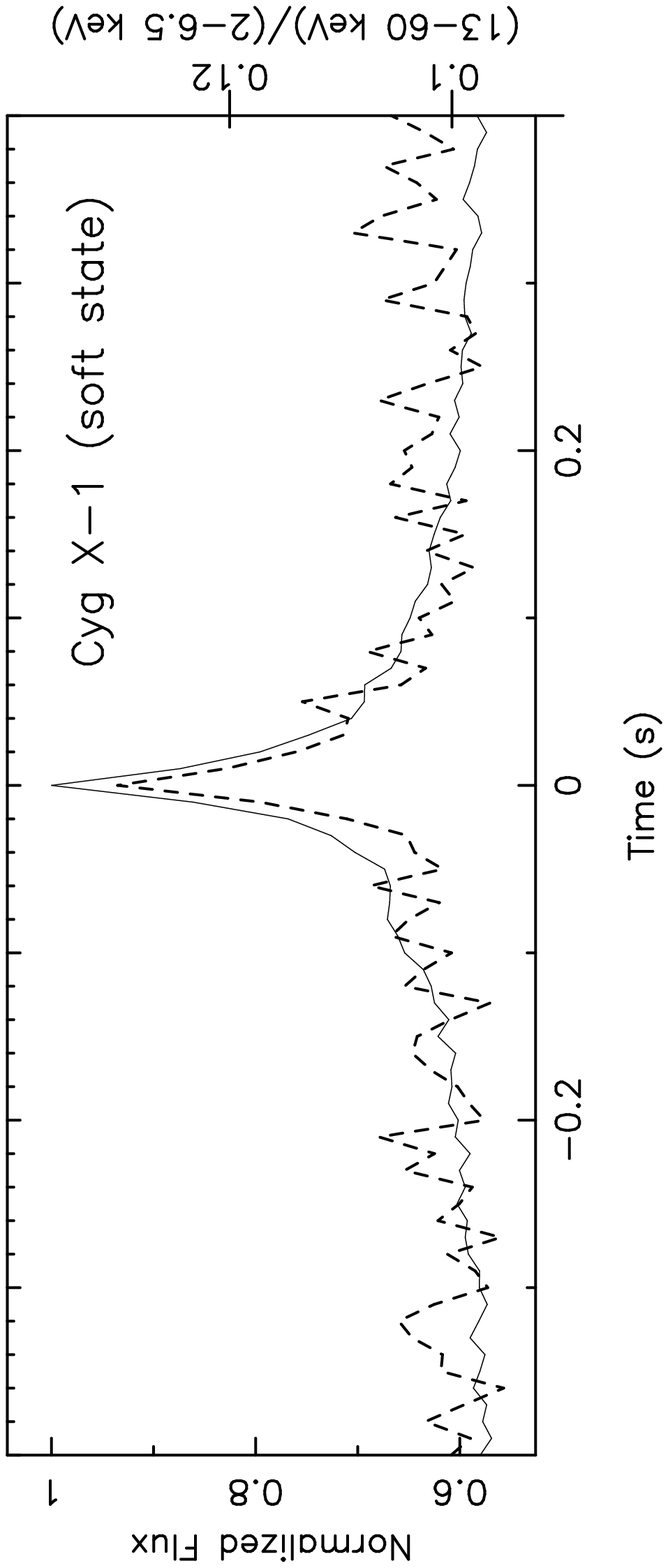}{20pt}{-90}{36}{50}{-50}{255}
\plotfiddle{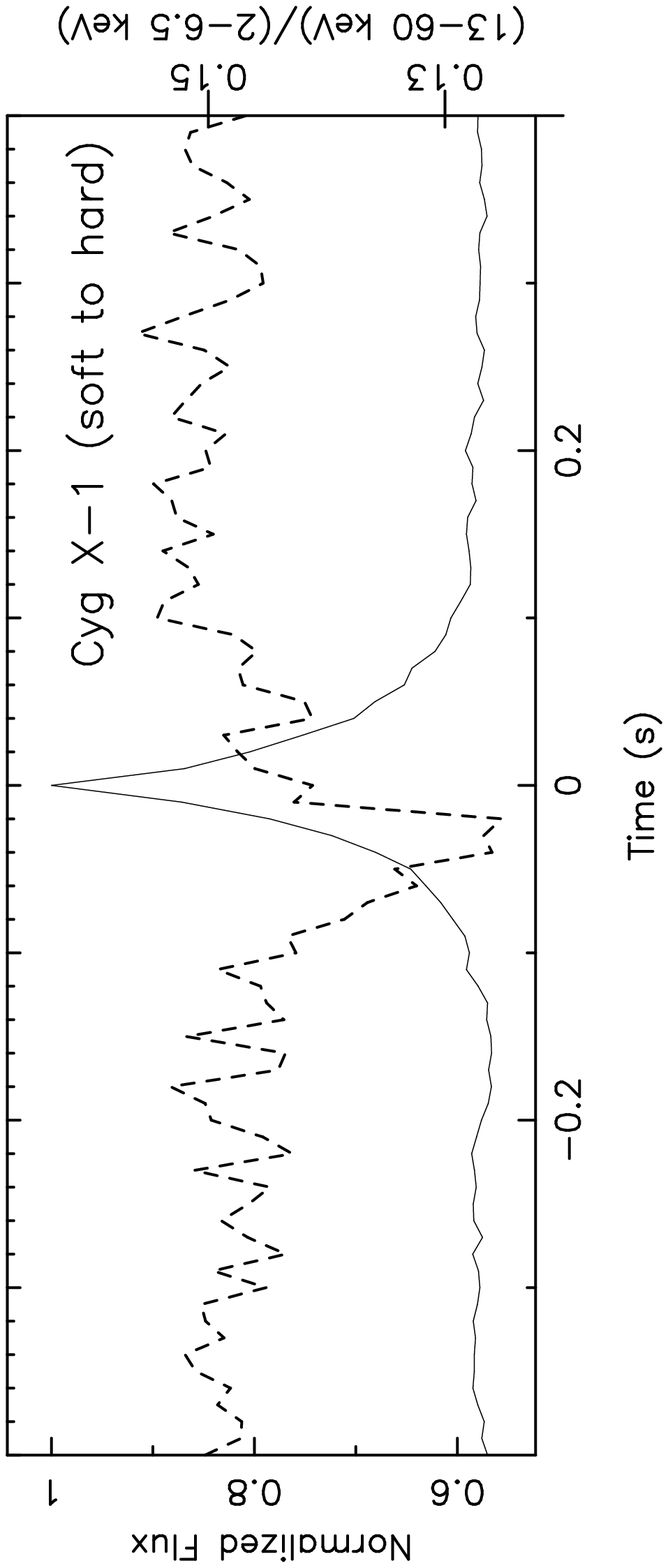}{20pt}{-90}{36}{50}{-300}{150}
\plotfiddle{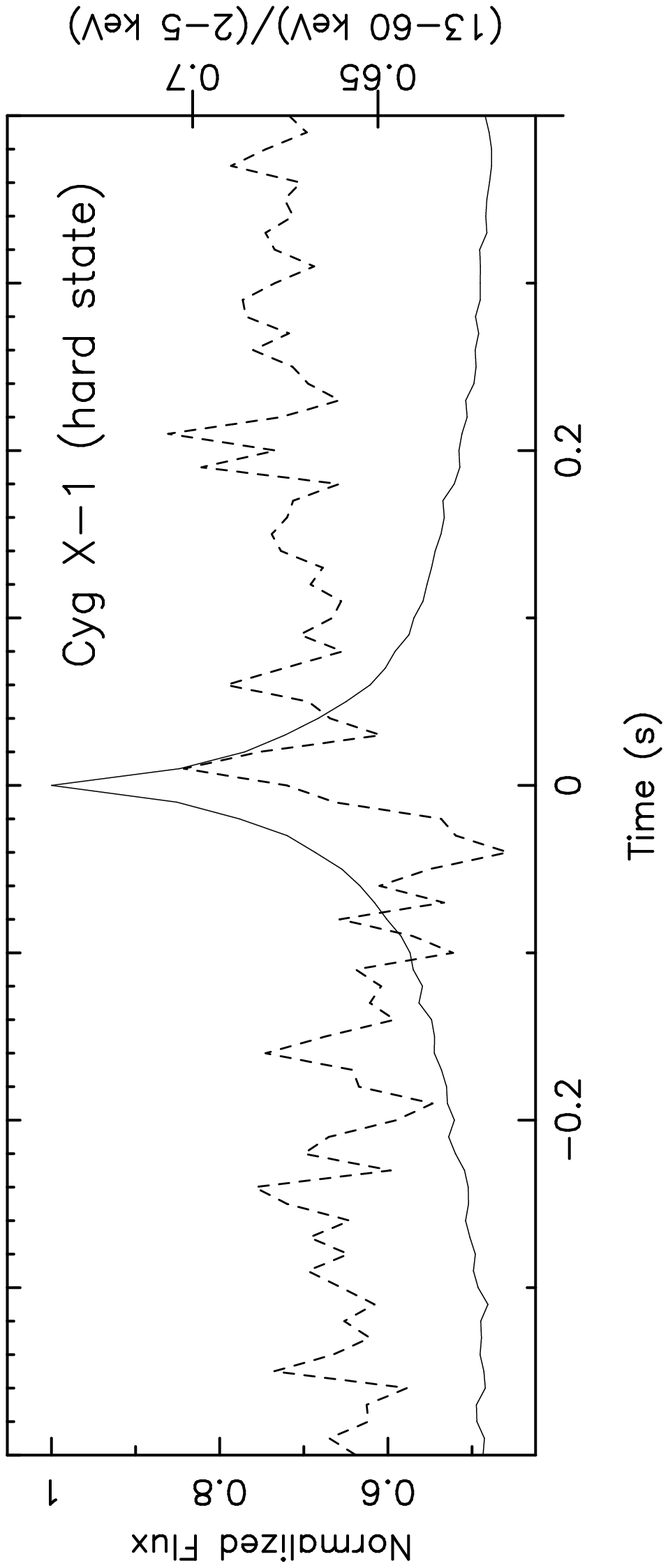}{20pt}{-90}{36}{50}{-50}{185}
\vspace{4cm}
\caption{The intensity and hardness ratio profiles of average shots in  different states
of Cyg X-1.  The solid lines are the superposed shot profiles with the peak height
being normalized to be unity (from  Fig.5 of \cite{fe98}). The dashed lines are profiles of
hardness ratio between the high and low energy band of average shots (from Fig.9 of
\cite{fe98}). (a) Upper-left panel: hard-to-soft transition.
(b) Upper-right panel: soft state. (c) Lower-left panel: soft-to-hard transition. 
(d) Lower-right panel: hard state. \label{fig6}}
\end{figure}

\clearpage
\hbox{}\vspace{2cm}
\begin{center}
Table 1. Average Correlation Coefficients between Hardness and Intensity   
\nopagebreak
\vspace{2mm}
\begin{tabular}{c | c | c c c c}
\cline{1-6} 
State & Start Time & \multicolumn{4}{c}{Time Scale (s)} \\
of Cyg X-1 & (Day of 1996) & $10^{-3}$ & $10^{-2}$ & $10^{-1}$ & 1 \\
\cline{1-6}
Hard & 142.7 & $-0.261\pm0.002$ & $-0.017\pm0.003$ & $-0.396\pm0.007$ & $-0.31\pm0.03$ \\
to   & 143.7 & $-0.194\pm0.001$ & $-0.011\pm0.002$ & $-0.430\pm0.007$ & $-0.59\pm0.02$ \\
Soft & 150.3 & $-0.349\pm0.005$ & $0.006\pm0.003$  & $-0.327\pm0.007$ & $-0.28\pm0.03$ \\
\cline{1-6}
     & 167.0 & $-0.235\pm0.008$ & $0.094\pm0.008$ & $0.17\pm0.02$ & $0.52\pm0.05$ \\
Soft & 168.3 & $-0.189\pm0.002$ & $0.101\pm0.002$ & $0.301\pm0.007$ & $0.58\pm0.02$ \\
     & 169.3 & $-0.174\pm0.005$ & $0.105\pm0.008$ & $0.33\pm0.02$ & $0.61\pm0.07$ \\
\cline{1-6}
Soft to& 223.3 & $-0.222\pm0.001$ & $-0.005\pm0.003$ & $-0.368\pm0.006$ & $-0.27\pm0.02$ \\
Hard & 224.6 &   $-0.371\pm0.007$ & $0.013\pm0.003$ & $-0.265\pm0.008$ & $-0.17\pm0.03$ \\
\cline{1-6}
     & 351.5 & $-0.392\pm0.002$ & $0.032\pm0.002$ & $0.051\pm0.007$ & $-0.18\pm0.02$ \\
Hard & 351.9 & $-0.546\pm0.002$ & $0.036\pm0.002$ & $0.067\pm0.008$ & $-0.08\pm0.02$ \\
     & 352.0 & $-0.613\pm0.002$ & $0.033\pm0.002$ & $0.062\pm0.006$ & $-0.11\pm0.02$ \\  
\cline{1-6}

\end{tabular}
\end{center}

\hbox{}\vspace{5mm} 
\begin{center}
Table 2. Average Correlation Coefficients between Hardness and Intensity  
\nopagebreak
\vspace{2mm}
\begin{tabular}{c| c| c c c }
\cline{1-5} 
State & Start Time & \multicolumn{3}{c}{Time Scale (s)} \\
of Cyg X-1 & (Day of 1996) &   10      & 20              & 50                 \\
\cline{1-5}
Hard & 142.7 & $0.06\pm0.09$ & $0.16\pm0.10$ & $0.33\pm0.27$  \\
to   & 143.7 & $-0.30\pm0.08$ & $-0.41\pm0.07$ & $-0.70\pm0.19$ \\
Soft & 150.3 & $-0.01\pm0.12$ & $-0.01\pm0.16$  & $-0.34\pm0.15$  \\
\cline{1-5}
     & 167.0 & $0.79\pm0.02$ & $0.95\pm0.08$ &     \\
Soft & 168.3 & $0.79\pm0.03$ & $0.85\pm0.03$ & $0.89\pm0.05$  \\
     & 169.3 & $0.53\pm0.27$ &  &  \\
\cline{1-5}
Soft to& 223.3 & $0.18\pm0.08$ & $0.04\pm0.10$ & $-0.27\pm0.11$  \\
Hard & 224.6 &   $0.12\pm0.09$ & $0.27\pm0.19$ & $0.38\pm0.02$  \\
\cline{1-5}
     & 351.5 & $-0.35\pm0.07$ & $-0.34\pm0.09$ & $0.07\pm0.13$  \\
Hard & 351.9 & $-0.43\pm0.06$ & $-0.36\pm0.11$ & $-0.33\pm0.16$  \\
     & 352.0 & $-0.35\pm0.05$ & $-0.29\pm0.09$ & $-0.14\pm0.16$  \\  
\cline{1-5}

\end{tabular}
\end{center}

\clearpage
\hbox{}\vspace{5mm}
\begin{center}
Table 3. FWHM of ACF of Low Energy Photons  
\nopagebreak
\vspace{2mm}
\begin{tabular}{c| c| c c c }
\cline{1-5} 
State & Start Time & \multicolumn{3}{c}{FWHM$_1/\Delta t$} \\
of Cyg X-1 & (Day of 1996) & $\Delta t=0.01$ s & $\Delta t=0.1$ s & $\Delta t=1$ s  \\
\cline{1-5}
Soft  & 168.3 & $0.410\pm0.002\pm0.012$&$1.63\pm0.01\pm0.05$ & $1.25\pm0.05\pm0.15$  \\
State & 169.3 & $0.469\pm0.002\pm0.007$ &$1.47\pm0.02\pm0.05$ &$1.08\pm0.07\pm0.13$ \\
\cline{1-5}
Transition &143.7&$0.548\pm0.002\pm0.012$&$1.58\pm0.01\pm0.04$&$0.71\pm0.02\pm0.09$  \\
State      &223.3&$0.503\pm0.002\pm0.007$&$1.42\pm0.01\pm0.04$&$0.77\pm0.04\pm0.18$ \\
\cline{1-5}
Hard  & 351.5 & $ 0.674\pm0.003\pm0.011$&$1.63\pm0.01\pm0.04$ & $1.41\pm0.05\pm0.14$\\
State & 351.9 & $ 0.691\pm0.003\pm0.013$ &$1.63\pm0.01\pm0.05$ &$1.31\pm0.05\pm0.17$\\
\cline{1-5}

\end{tabular}

\end{center}

\hbox{}\vspace{5mm}
\begin{center}
Table 4. ACF Width Ratio of High Energy Photons to Low Energy Photons      
\nopagebreak
\vspace{2mm}
\begin{tabular}{c| c| c c c }
\cline{1-5} 
State & Start Time & \multicolumn{3}{c}{$FWHM_3/FWHM_1$} \\
of Cyg X-1 & (Day of 1996) & $\Delta t=0.01$ s & $\Delta t=0.1$ s & $\Delta t=1$ s  \\
\cline{1-5}
Soft  & 168.3 & $2.09\pm0.01\pm0.11$&$0.87\pm0.01\pm0.05$&$1.11\pm0.07\pm0.12$  \\
State & 169.3 & $2.15\pm0.01\pm0.06$&$0.75\pm0.01\pm0.05$&$0.91\pm0.07\pm0.16$ \\
\cline{1-5}
Transition &143.7&$1.440\pm0.008\pm0.019$&$0.747\pm0.009\pm0.016$&$0.87\pm0.05\pm0.19$  \\
State      &223.3&$1.580\pm0.009\pm0.015$&$0.785\pm0.009\pm0.021$&$1.05\pm0.08\pm0.24$ \\
\cline{1-5}
Hard  & 351.5 & $1.194\pm0.006\pm0.022$&$0.94\pm0.01\pm0.03$ &$0.97\pm0.05\pm0.10$ \\
State & 351.9 & $1.173\pm0.007\pm0.014$&$0.95\pm0.01\pm0.03$ &$0.97\pm0.05\pm0.14$ \\
\cline{1-5}

\end{tabular}

\end{center}

\vspace{5mm}
\begin{center}
Table 5. Time Delay  of High Energy Photons  to Low Energy Photons
\nopagebreak
\vspace{2mm}
\begin{tabular}{c| c| c c c }
\cline{1-5} 
State & Start Time & \multicolumn{3}{c}{Time Delay (ms) }\\
of Cyg X-1 & (Day of 1996) & $\Delta t=0.01$ s & $\Delta t=0.1$ s & $\Delta t=1$ s  \\
\cline{1-5}
Soft  & 168.3 & $0.700\pm0.005\pm0.024$&$17.1\pm0.5\pm1.8$&$28.3\pm2.2\pm9.2$  \\
State & 169.3 & $0.817\pm0.009\pm0.047$&$24.0\pm0.8\pm1.6$ & $37.9\pm5.0\pm52.3$ \\
\cline{1-5}
Transition &143.7& $0.696\pm0.004\pm0.015$&$23.3\pm0.5\pm1.7$ & $23.2\pm2.3\pm24.2$  \\
State      &223.3 &$0.688\pm0.004\pm0.009$&$23.1\pm0.4\pm1.4$ & $32.1\pm2.7\pm14.0$ \\
\cline{1-5}
Hard  & 351.5 & $0.749\pm0.006\pm0.014$&$16.4\pm0.4\pm0.8$&$22.1\pm1.9\pm5.5$  \\
State & 351.9 & $0.751\pm0.006\pm0.016$&$16.2\pm0.4\pm1.5$&$22.3\pm2.2\pm7.4$ \\
\cline{1-5}

\end{tabular}
\end{center}

\end{document}